\documentclass[iop]{emulateapj}
\usepackage{hyperref}
\usepackage{amsmath}
\usepackage{algorithm}
\usepackage{amsfonts}
\usepackage{mathrsfs}
\usepackage{bm}
\usepackage{rotating}
\usepackage{color}
\usepackage{graphicx}
\usepackage{subfigure}
\usepackage{lineno}

\def\cha{\textit{Chandra}}
\def\XMM{{XMM-{\it Newton}}}
\def\NuSTAR{{\it NuSTAR}}
\def\bat{{{\it Swift}-BAT}}
\def \XSPEC {{\tt XSPEC}}
\def \pexrav {{\tt pexrav}}
\def \MYTorus {{\tt MYTorus}}
\def \borus {{\tt borus02}}
\def \bntorus {{\tt BNtorus}}

\begin{document}
\title{Compton-thick AGN in the \NuSTAR\ era IV: A deep \NuSTAR\ and \XMM\ view of the candidate Compton thick AGN in ESO 116-G018}
\author{X. Zhao\altaffilmark{1}, S. Marchesi\altaffilmark{1}, M. Ajello\altaffilmark{1}}

\altaffiltext{1}{Department of Physics \& Astronomy, Clemson University, Clemson, SC 29634, USA}

\begin{abstract}
We present the 2--78\,keV spectral analysis of the deep \NuSTAR\ and \XMM\ observation of a nearby Seyfert 2 galaxy, ESO 116-G018, which is selected as a candidate Compton-thick (CT-) active galactic nucleus (AGN) based on a previous \cha-\bat\ study. Through our analysis, the source is for the first time confirmed to be a CT-AGN at a $>$3\,$\sigma$ confidence level, with the ``line-of-sight'' column density N$_{\rm H,Z}$ = [2.46--2.76] $\times$ $10^{24}$\,cm$^{-2}$. The ``global average'' column density of the obscuring torus is N$_{\rm H,S}$ = [0.46--0.62] $\times$ $10^{24}$\,cm$^{-2}$, which suggests a clumpy, rather than uniform, distribution of the obscuring material surrounding the accreting supermassive black hole. The excellent-quality data given by the combined \NuSTAR\ and \XMM\ observations enable us to produce a strong constraint on the covering factor of the torus of ESO 116-G018, which is found to be $f_c$ = [0.13-0.15]. We also estimate the bolometric luminosity from the broad-band X-ray spectrum, being L$\rm _{bol}$ = [2.57--3.41] $\times$ 10$^{44}$\,erg\,s$^{-1}$.
\end{abstract}

\keywords{galaxies: active -- galaxies: nuclei -- galaxies: individual (ESO 116-G018) -- X-rays: galaxies}

%
%
\section{Introduction}\label{sec:intro} 
The intrinsic emission from an accreting supermassive black hole (SMBH), i.e., the center of active galactic nuclei (AGNs), is commonly believed to be, at least partly, obscured by the circumnuclear matter. Especially, AGN are classified as Compton-thick (CT-) AGNs when the column density of the obscured matter is N$\rm_H$ $\ge$ $\sigma_T^{-1}$ $\sim$10$^{24}$\,cm$^{-2}$, where $\sigma_T$ is the Thomson cross section.
Knowing the distribution of the absorbing column density is not only important to understand the physics of the accreting SMBHs, but is also essential to properly model the cosmic X-ray background (CXB), i.e., the diffused X-ray emission observed between 0.5\,keV and 300\,keV, which is believed to be mainly produced by both obscured and unobscured AGNs. While most of the CXB emission below 10\,keV has been resolved thanks to \cha\ and \XMM\ \citep[see, e.g.,][]{Worsley05,Hickox06}, only $\sim$35\% of the CXB emission at its peak \citep[$\sim$30\,keV,][]{Ajello08} has been resolved, mostly by different \NuSTAR\ surveys \citep{aird15,harrison15}. In this energy range, CT-AGNs are expected to be numerous \citep[up to 50\% of the overall population of Seyfert 2 galaxies, see, e.g.,][]{risaliti1999}. Different CXB synthesis models predict that the fraction of CT-AGN should be $\sim$20\%--30\% \citep[][]{Alexander03,Gandhi03,gilli07,Treister09,Ueda14}. However, as of today CT-AGNs have never been detected in large numbers, e.g., their observed fraction in the local universe is $\sim$5--10\% \citep[see, e.g.][]{Burlon11,Ricci15} in the X-rays and is $\sim$12\% when performing a multi-wavelength search \citep[][]{Goulding11}. 

Due to the heavy obscuration, CT-AGNs are difficult to detect below $\sim$10\,keV in the local universe \citep[see, e.g.,][]{gilli07,koss2016}, since the overall X-ray emission in these objects is suppressed below 10\,keV and  dominated by the Compton hump at $\sim$20--40\,keV. CT-AGNs at redshift $z>$1 can instead be well studied using one of the several facilities sampling the $\sim$0.3--10\,keV energy range, such as \textit{Swift}-XRT, \cha, \XMM\ and \textit{Suzaku} \citep[see, e.g.,][]{Georgantopoulos2013,Buchner2015,Lanzuisi2105}: the Compton hump of high-$z$ sources is redshifted in the energy range covered by these instruments. However, for sources in the local universe ($z$ $<$0.1), the proper characterization of heavily obscured AGN requires an X-ray telescope sensitive above 10\,keV. Thanks to the launch of \textit{Nuclear Spectroscopic Telescope Array} \citep[hereafter, \NuSTAR,][]{harrison}, which provides a two orders of magnitude better sensitivity than previous telescopes \citep[e.g., \textit{INTEGRAL} and \bat;][]{Winkler2003,Barthelmy2005} at $\sim$10--50\,keV, one can study the physical and geometrical properties of heavily obscured AGN with unprecedented accuracy \citep[see, e.g.][]{Balokovic14,puccetti14,Annuar15,Stefano2017,Marchesi2018,Ursini18}. 
To properly constrain both the torus column density and the AGN photon index in heavily obscured sources, however, one needs to combine the excellent \NuSTAR\ effective area at energies $>$10\,keV with a soft X-ray instrument, which covers the 0.5--10\,keV energy range.
Among these, \XMM\ is the best one in terms of both effective area in the 0.3--10\,keV energy range and spectral energy resolution (150\,eV, $\sim$2.5 better than \NuSTAR, which has $\Delta$E = 400\,eV) at the energy of the Fe K$\alpha$ line (the signature of obscured AGN at E=6.4\,keV). 

The obscuration observed in AGNs across the electromagnetic spectrum, from X-ray to infrared, is usually explained with a pc-scale, torus-like structure of dust and gas \citep[see, e.g.,][]{Natureastro2017}. Consequently, several tori models, based on Monte Carlo simulations, have been developed to characterize the X-ray spectra of CT-AGNs in the past two decades \citep{Matt1994,Shinya09,MYTorus2009,BNtorus,Liu14,Furui16,Borus}. All these models assume a continuous distribution of the obscuring material, but with a different assumption on the geometry of the torus. In particular, in the models proposed by \citet{Shinya09}, \citet{BNtorus} and \citet{Borus}, the half opening angle of the torus, i.e., the torus covering factor, is a free parameter, thus allowing to put constraints on the toroidal geometry. Given the intrinsic complexity of these models, and the multiple free parameters involved, applying them in full capability requires high-quality X-ray spectra, with excellent statistics on a wide energy range, i.e., between 1 and 100\,keV: at the present day, similar requirements can be satisfied only by a joint \NuSTAR\ and \XMM\ observation.
The AGN emission can also be observed at infrared wavelengths, where part of the intrinsic accretion disk optical-UV emission is absorbed by the dust in the ``torus-like'' structure and then re-emitted in the infrared. Thus, the fraction of the luminosity of the torus with respect to the AGN bolometric luminosity ($L_{\rm tor}$/$L_{\rm AGN}$) can be used as a proxy of the torus covering factor \citep[see; e.g.][]{Stalevski16}. Indeed, in addition to the previously mentioned X-ray models, theory and models on the nature of the obscuration from an infrared perspective have also been developed \citep{Krolik1988,Jaffe04,Tristram07,Nenkova08,Honig10,Stalevski12}.

In this work, we present the results of a deep, 50\,ks combined \NuSTAR\ and \XMM\ observation of ESO 116-G018, a nearby Seyfert 2 galaxy and a candidate CT-AGN. The paper is organized as follows: in Section \ref{sec:Observ} , we report the \NuSTAR\ and \XMM\ data reduction and spectral extraction process; in Section \ref{sec:spectral}, we describe the different models which are used to fit the broadband X-ray spectra, and the results of the spectral analysis using above models; in Section \ref{discussion}, we compare our results with those already existent in the literature, and discuss the constraints on the physical properties of ESO 116-G018, e.g., the equivalent width of the iron K$\alpha$ line and the intrinsic luminosity, and the geometry, i.e., covering factor, of the obscuring ``torus-like'' structure.
All reported uncertainties on spectral parameters are at 90\% confidence level, if not otherwise stated. Standard cosmological constants are adopted as follows: $<H_0>$ = 70 km s$^{-1}$ Mpc$^{-1}$, $<q_0>$ = 0.0 and $<\Lambda>$ = 0.73.
%
%
\section{Observation and Data Analysis}\label{sec:Observ}
ESO 116-G018 \citep[$z$ $\sim$0.0185, $d$ $\sim$80\,Mpc,][]{Grijp1992} is a Seyfert 2 galaxy which is detected in the 100-month BAT catalog (Segreto et al. 2019 in prep.), a catalog of $\sim$1000 AGNs detected by \textit{Swift}-BAT in the 15--150\,keV band. 

ESO 116-G018 was first selected as a candidate CT-AGN by \citet{marchesi2017APJ} using the selection technique described in the paper and then was targeted with a 10\,ks follow-up observation with \cha. The joint \cha--\bat\ spectral fit allowed us to obtain a first measurement of the source ``line-of-sight" column density, which is N$_{\rm H,Z}$ =  0.95$_{-0.40}^{+0.46}$ $\times$ $10^{24}$\,cm$^{-2}$. The low-quality of the \cha\ spectrum ($\sim$50 net counts in the 0.5--7\,keV band) prevented us to properly characterize ESO 116-G018, or even confirm or reject its Compton-thick origin at a $>$3\,$\sigma$ confidence level. Therefore, to further investigate this new candidate CT-AGN, as well as another one, NGC 1358, which we analyzed in a companion paper (Zhao et al. 2018 submitted), we proposed for a simultaneous \NuSTAR\ (45\,ks) and \XMM\ (58\,ks) follow-up observation, which was accepted in \NuSTAR\ Cycle 3 (proposal ID 3258, PI: Marchesi). We report a summary of the observations in Table \ref{tab:obs_summary}.

\begin{table*}
\center
\caption{Summary of \NuSTAR\ and \XMM\ observations.}
\vspace{.1cm}
  \begin{tabular}{cccccc}
       \hline
       \hline
    Instrument&Sequence&Start Time&End Time&Exposure Time&Count Rate\tablenotemark{a}\\ 
    &ObsID&(UTC)& (UTC)&(ks)&$10^{-2}$counts s$^{-1}$\\
    \hline
    \NuSTAR&60301027002&2017-11-01T18:56:09&2017-11-02T20:46:09&45& 1.24$\pm$0.06  1.31$\pm$0.06 \\
    \XMM&0795680201&2017-11-01T19:19:57&2017-11-02T11:32:32&58&0.33$\pm$0.02 0.44$\pm$0.03 1.58$\pm$0.06\\
       \hline
\end{tabular}
\par
\vspace{.2cm}
\tablenotemark{a}{The reported \NuSTAR\ count rates are those of the FPMA and FPMB modules between 3--78\,keV, respectively. The reported \XMM\ count rates are those the MOS1, MOS2 and pn modules between 2--10\,keV, respectively.}\label{tab:obs_summary}
\end{table*}

\subsection{\NuSTAR\ Observation}
ESO 116-G018 was observed by \NuSTAR\ on 2017 November 1--2 (ObsID 60301027002). The observation took place in a 95.5\,ks time-span and was divided into 15 ($\sim$3\,ks) intervals. The gaps in the observation correspond to the periods of time in which the target was occulted by the Earth.
The \NuSTAR\ data is derived from both focal plane modules, FPMA and FPMB. The raw files are calibrated, cleaned and screened using the \NuSTAR\ \texttt{nupipeline} script version 0.4.5. The \NuSTAR\ calibration database (CALDB) used in this work is version 20171002. The ARF, RMF and light-curve files are obtained using the \texttt{nuproducts} script. 
For both modules, the source spectrum is extracted from a 30$^{\prime\prime}$ circular region, corresponding to $\approx$50\% of the encircled energy fraction (EEF) at 10\,keV, centered on the source optical position. We then extract a background spectrum for each module, choosing a 30$^{\prime\prime}$ circular region located nearby the outer edges of the field of view, to avoid contamination from the source and no flares are found in the background light curves. The \NuSTAR\ spectra are grouped with a minimum of 15 counts per bin using the HEAsoft task \texttt{grppha}. 

\subsection{\XMM\ Observation} 
The \XMM\ observation was taken quasi-simultaneously to the \NuSTAR\ one with the EPIC CCD cameras \citep[pn;][]{pn} and two MOS cameras \citep[][]{MOS}: the \XMM\ observation started at the same time, but ended $\sim$9 hours before the \NuSTAR\ one. We reduced the \XMM\ data using the Science Analysis System \citep[SAS;][]{SAS} version 16.1.0. The source spectra are extracted from a 15$^{\prime\prime}$, corresponding to $\approx$70\% of the encircled energy fraction (EEF) at 1.5\,keV, circular region, while the background spectra are obtained from an 80$^{\prime\prime}$ circle located nearby the source. We visually inspected the \XMM\ image to avoid contamination to the background from sources nearby ESO 116-G018.

\subsection{Variability}\label{variability}
\begin{figure*} 
\begin{minipage}[b]{.5\textwidth}
\centering
\includegraphics[width=1\textwidth]{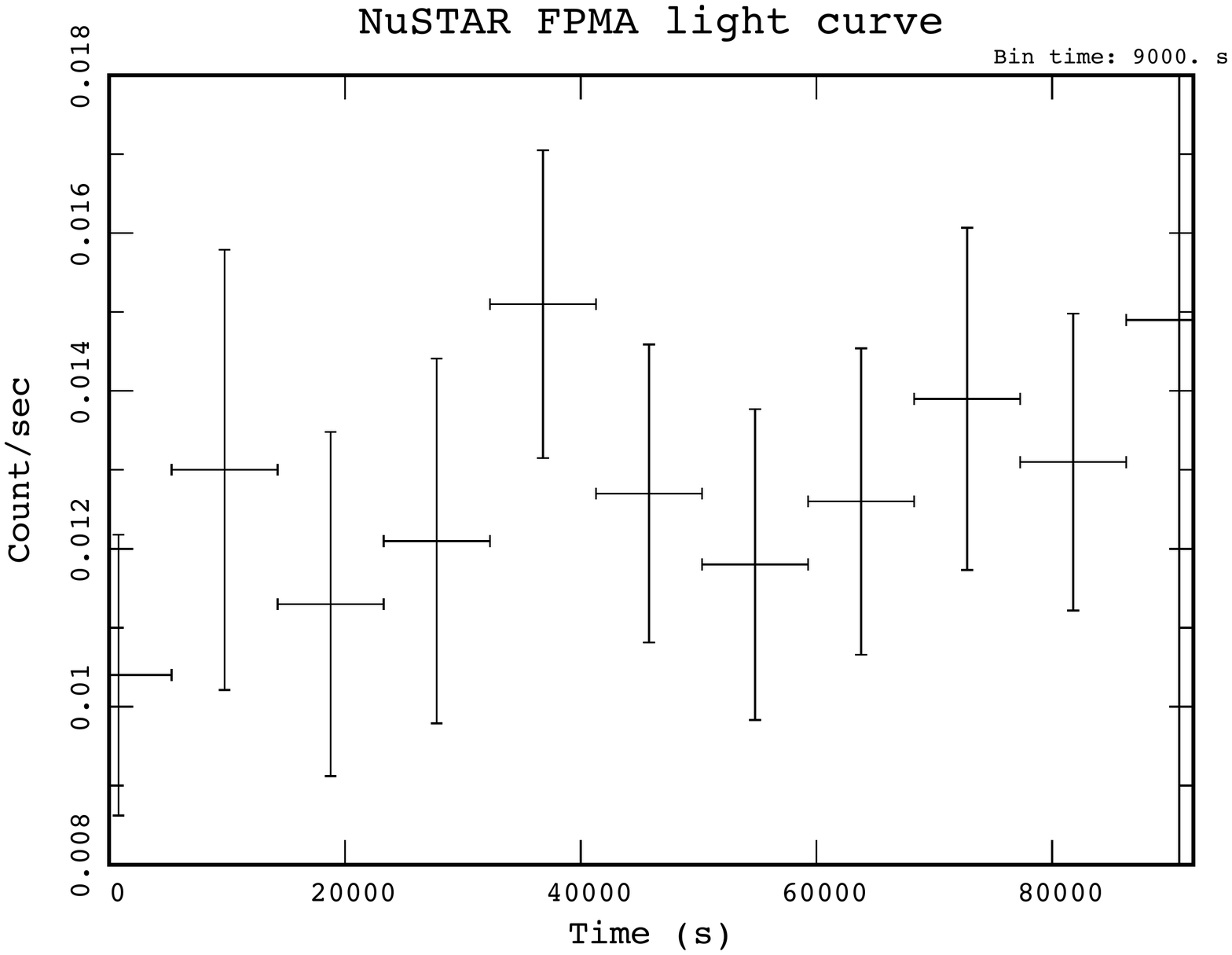}
\end{minipage}
\begin{minipage}[b]{.5\textwidth}
\centering
\includegraphics[width=1\textwidth]{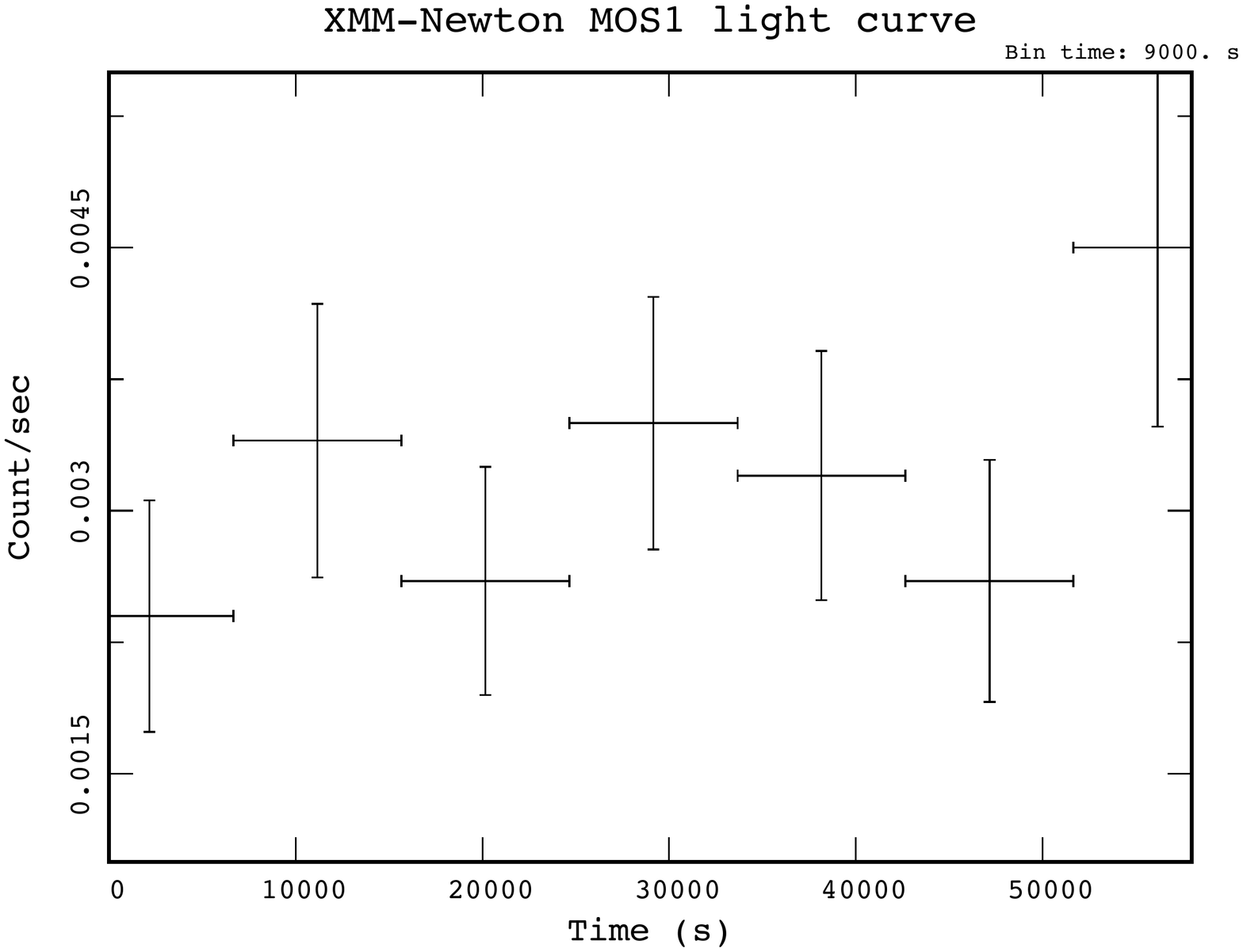}
\end{minipage}
\caption{Background subtracted light curves of \NuSTAR\ module FPMA (left panel) \XMM\ module MOS1 (right panel). The bin time is equal to 9 ks for both light curves. The average count rates are $r_{\rm FPMA}$ = 1.3$\pm{0.3}$ $\times$ 10$^{-2}$\,cts s$^{-1}$ for FPMA and r$\rm_{MOS1}$ = 3.2$\pm{0.7}$ $\times$ 10$^{-3}$\,cts s$^{-1}$ for MOS1, respectively.}
\label{fig:lightcurve}
\end{figure*}

When visually inspecting the light curves of both \NuSTAR\ (3--78\,keV) and \XMM\ (2--10\,keV) of ESO 116-G018, we find no obvious evidence of variability during the observations. The background subtracted light curves of \NuSTAR\ module FPMA and \XMM\ EPIC MOS1 are presented in Figure \ref{fig:lightcurve}. We further analyze the two light curves by fitting them with a constant, $r$, which corresponds to the average count rate: we use the $\chi^2$ test to check for any statistical evidence of variability. The best-fit average count rate is $r_{\rm FPMA}$ = 1.3$\pm{0.3}$ $\times$ 10$^{-2}$\,cts s$^{-1}$ for the \NuSTAR\ module FPMA; the $\chi^2$ for the fit is $\chi^2_{\rm FPMA}$ = 4.4 with 10 degrees of freedom, while the light curve would be different from a constant at a the $>$99\% confidence level if $\chi^2$ $>$23.2 for 10 degrees of freedom. The best-fit average count rate of \XMM\ EPIC MOS1 is $r_{\rm MOS1}$ = 3.2$\pm{0.7}$ $\times$ 10$^{-3}$\,cts s$^{-1}$; the $\chi^2$ for the fit is $\chi^2_{\rm MOS1}$ = 4.9, while the light curve would be different from a constant at a the $>$99\% confidence level if $\chi^2$ $>$16.8 for 6 degrees of freedom. 

Based on the fit statistics given above, there is no significant variability in both the \NuSTAR\ and the \XMM\ light curve of ESO 116-G018.

%
%
\section{Spectral Modeling Results}
\label{sec:spectral}
We performed the spectral fit of ESO 116-G018 using \XSPEC\ v12.9.1 \citep{Arnaud1996} and the $\chi^2$ statistic. The photoelectric cross section for all absorption components used are those from \cite{Verner1996}. The element abundance is from \citet{Anders1989} and metal abundance is fixed to Solar. The Galactic absorption column density is N$\rm _{H,Gal}$ = $3.1\times 10^{20}$\,cm$^{-2}$ \citep{Kalberla05}. The source redshift is fixed at $z$ = 0.0185.

Following a standard approach in analyzing heavily obscured AGN, we begin our spectral modeling using the phenomenological model. We report in Table \ref{Table:best-fit}  the results of the joint \NuSTAR--\XMM\ spectral fitted using the different models which will be discussed in the following sections.

\subsection{Phenomenological Model}\label{sec:pheno}
\label{sec:model_abs}
We first fit the spectra with a phenomenological model composed of an absorbed power law with photon index $\Gamma$. The absorption caused by the obscuring gas and dust surrounding the accreting SMBH is modeled by \texttt{zphabs}, while the Galactic absorption is modeled by \texttt{phabs}. We also add to the model a Gaussian (\texttt{zgauss}) to characterize the prominent Fe K$\alpha$ line typically observed in heavily obscured AGNs. We fix the center of the Gaussian at 6.4\,keV and fix the line width $\sigma$ to 50\,eV, assuming the line to be narrow, to minimize the number of free parameters: nonetheless, no significant improvement is found when leaving the line width free to vary. Below $\sim$5\,keV, the spectrum is dominated by the fraction \citep[usually less than 5--10\%, see, e.g.,][]{Marchesi2018} of emission from the intrinsic X-ray continuum, which is not intercepted by the torus on the ``line-of-sight'', and/or the intrinsic emission is deflected, rather than absorbed by the obscuring material into the ``line-of-sight''.
This scattered component is modeled by an unabsorbed power law having photon index $\Gamma_2$ = $\Gamma$: the fractional intensity with respect to the intrinsic emission, $f_s$, is modeled by a constant ($constant_2$).

Finally, the cross-calibration between \NuSTAR\ and \XMM\ is modeled by another constant ($constant_1$), noted as $C_{NuS/XMM}$. We also assume that there is no flux offset between different modules of the same instrument. The phenomenological model (Model A), in \XSPEC\ nomenclature, is thus:
\begin{equation}
\label{eq:powerlaw}
\begin{aligned}
Model A = &constant_1*phabs*(zphabs*zpowerlw\\
&+zgauss+constant_2*zpowerlw)
\end{aligned}
\end{equation}

\begin{figure}[htpb]
\centering
\includegraphics[width=.5\textwidth]{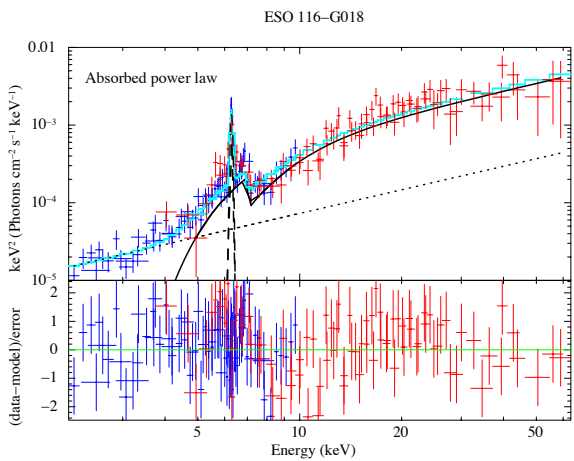} 
 \caption{Unfolded \XMM\ and \NuSTAR\ spectrum of ESO 116-G018 fitted with model A (top). The \XMM\ data are plotted in blue, while the \NuSTAR\ data are plotted in red. The best-fit models prediction is plotted as a cyan solid line. The single components of the model are plotted in black with different line styles, i.e., the absorbed intrinsic continuum as a solid line, the Fe K$\alpha$ line as a dashed line and the scattered component as a dotted line.}
 \label{fig:simple}
 \end{figure}

The best-fit results of model A is reported in Table \ref{Table:best-fit} and Fig.~\ref{fig:simple} shows the best-fit of the spectra of ESO 116-G018 fitted with model A. The best-fit intrinsic photon index is $\Gamma$ = 0.99$_{-0.12}^{+0.13}$ and the column density of the obscuring material along our ``line-of-sight'' is N$_{\rm H}$ = 0.62$_{-0.12}^{+0.12}$ $10^{24}$\,cm$^{-2}$. Although the statistics ($\chi^2_\nu$ = $\chi^2$/degrees of freedom, d.o.f. hereafter, = 166/162 = 1.02) of the phenomenological model is acceptable, the best-fit photon index, $\Gamma$ = 0.99$_{-0.12}^{+0.13}$, is not physically plausible \citep[typical AGNs have photon indices within the range $\Gamma$ = 1.4--2.6; see, e.g.,][]{MYTorus2009}. 
This is not an unexpected result, since the complexity of the spectral shape of a heavily obscured AGN cannot be properly treated by a standard absorption component alone such as \texttt{zphabs}: such a model cannot, for example, properly model the shape of the reprocessed component known as ``Compton hump" observed at energies $E$ $\sim$10--40\,keV. 

In the next section, we, therefore, fit the data with physically motivated models, i.e.,\pexrav\ \citep{pexrav}, \MYTorus\ \citep{MYTorus2009} and \borus\ \citep{Borus}, which are suitable to characterize heavily obscured AGNs with high-quality X-ray spectra.

\subsection{Physical Models}\label{sec:physical}

\subsubsection{Absorbed power-law with reflection component}\label{sec:pexrav}
\pexrav\ has historically been used to model heavily obscured AGN spectra where the observed emission is dominated by the photons reprocessed and upscattered by the obscuring material. 
\pexrav\ models a power law spectrum with an exponential cut off reflected from a slab of neutral material. We first test the \pexrav\ model utilized as a pure reflector by setting the reflection scaling factor to be R = -1, assuming the ``line-of-sight'' is heavily obscured (e.g., when N$\rm_H$ $\ge$10$^{25}$\,cm$^{-2}$) such that the observed spectrum is entirely contributed by the reflection from the back-side of the obscuring matter. The photon index is $\Gamma$ = 1.57$_{-0.09}^{+0.09}$ with the reduced $\chi^2$ to be $\chi^2_\nu$ = 167/163 = 1.02. Although the pure reflection component fits the reprocessed emission at 10--40\,keV well, it fails to describe the soft X-ray part at $E$ $<$10\,keV. Therefore, we add an absorbed power law to model the ``line-of-sight'' continuum following the method adopted in \citet{Ricci11}. 

In the \XSPEC\ nomenclature, our \pexrav\ model is written as follows:
\begin{equation}\label{eq:pexrav}
\begin{aligned}
Model B=&constant_1*phabs*(zphabs*zpowerlw+\\
&pexrav+zgauss+constant_2*zpowerlw)
\end{aligned}
\end{equation}
where all the components other than \pexrav\ are those already described in the previous section. While in \pexrav, the inclination angle $i$, i.e., the angle between the axis of the AGN (normal to the disk) and the observer line of sight, is fixed to $i$ = 60$^{\circ}$ (cos $i$ = 0.5). We do not find any significant variation in the spectral fit results when adopting other two inclination angle values, i.e., $i$ = 87$^{\circ}$ and 18$^{\circ}$ (cos $i$ = 0.05 and 0.95). The cut-off energy is fixed at $E_{cut}$ = 500\,keV to be consistent with the \MYTorus\ model, which will be discussed in detail in the following section: no significant improvement is found when we leave the cut-off energy free to vary. Finally, the reflection scaling factor $R$ is set to be less than 0 (i.e., the model describes only the reprocessed component) and is free to vary.  
 
\begin{figure*}
\begin{minipage}[b]{.5\textwidth}
\centering
\includegraphics[width=\textwidth]{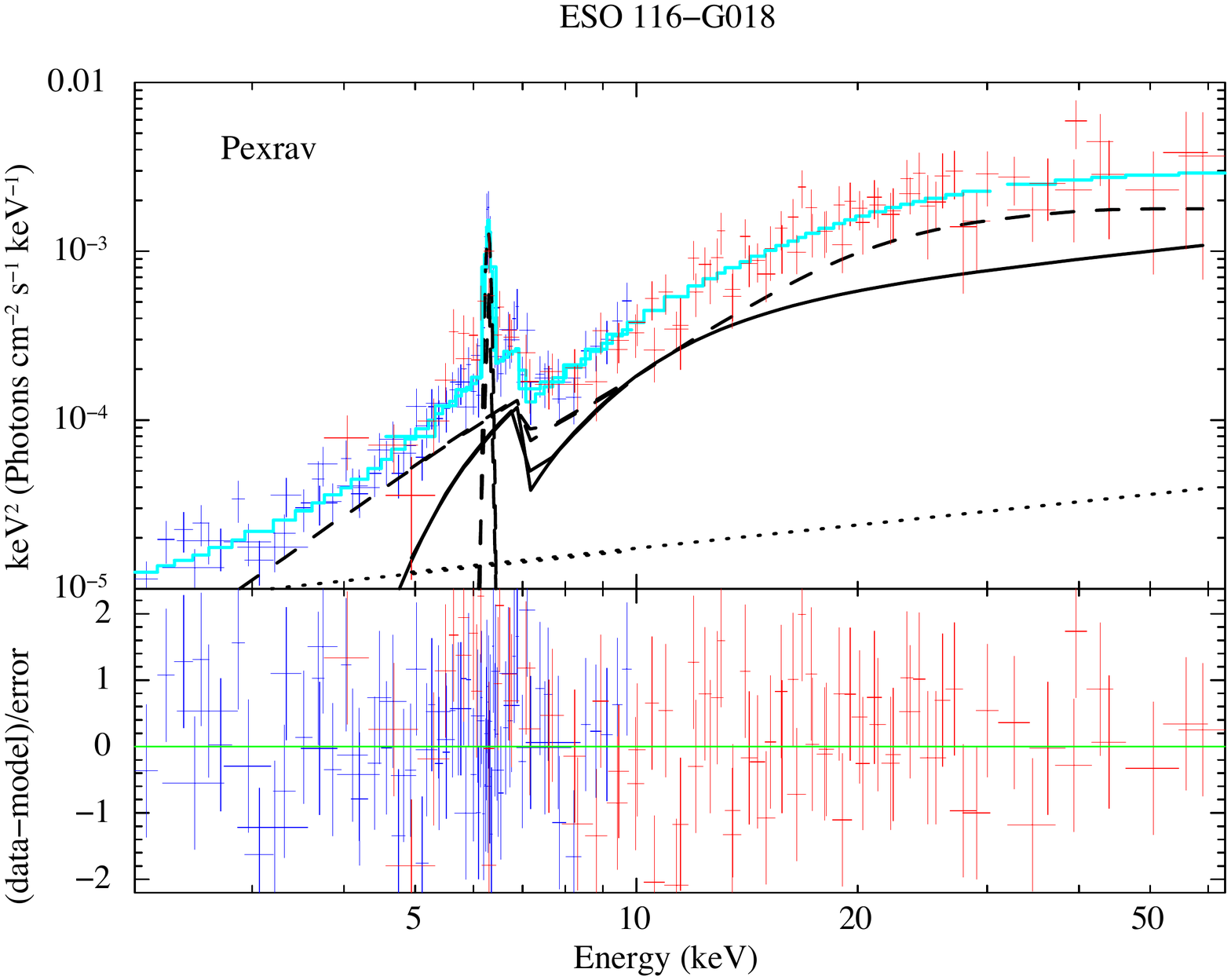}
\end{minipage}
\begin{minipage}[b]{.5\textwidth}
\centering
\includegraphics[width=\textwidth]{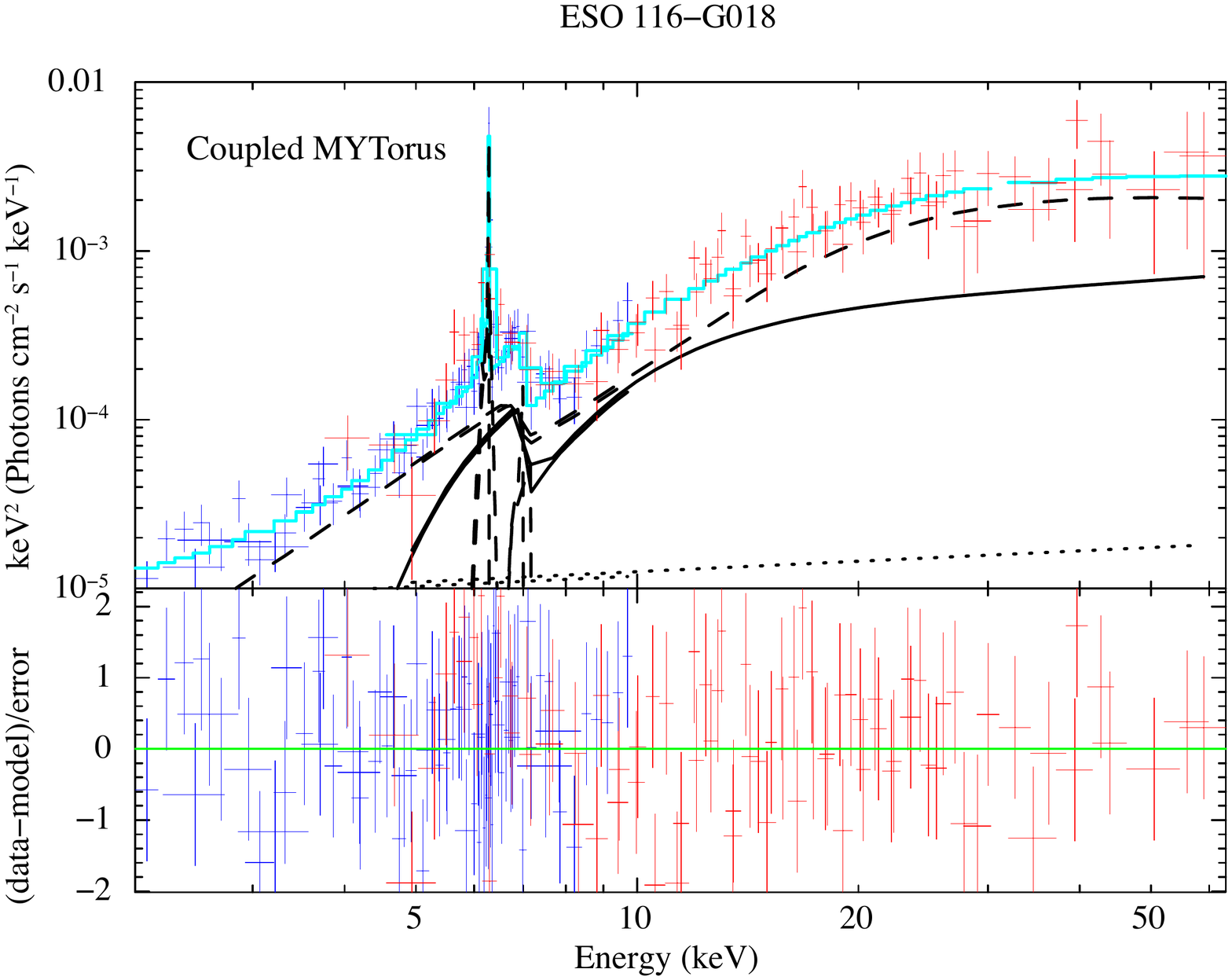}
\end{minipage}
\caption{Unfolded \XMM\ and \NuSTAR\ spectra of ESO 116-G018 fitted with model B (left panel) and `coupled' \MYTorus\ model (right panel). The \XMM\ data are plotted in blue, while the \NuSTAR\ data are plotted in red. The best-fit models prediction is plotted as a cyan solid line. The single components of the model are plotted in black with different line styles, i.e., the absorbed intrinsic continuum as a solid line, the reprocessed component and Fe K$\alpha$ line as a dashed line and the scattered component as a dotted line.}
 \label{fig:pex_coup}
 \end{figure*}

We show in Fig.~\ref{fig:pex_coup} the best-fit of the joint \XMM--\NuSTAR\ spectra obtained using Model B. The best-fit photon index is $\Gamma$ = 1.54$_{-0.19}^{+0.18}$ and the ``line-of-sight" column density is N$_{\rm H,Z}$ = 0.88$_{-0.26}^{+0.32}$ $\times$ $10^{24}$\,cm$^{-2}$. It is worth noting that, although the \pexrav\ result, with the reduced $\chi^2$ being $\chi^2_\nu$ = $\chi^2$/d.o.f. = 145/161 = 0.90, is  improved with respect to the one of Model A, different components of the spectrum, e.g., the iron line (modeled by a Gaussian), are not treated in a self-consistent way.

In summary, according to both model A and model B, ESO 116-G018 is heavily obscured but not Compton-thick. However, in model A, the photon index $\Gamma$ is significantly harder than the one of a typical obscured AGN; furthermore, the fraction of scattered emission is slightly larger than the typical 1--10\% value in both models, being $f_s$ $\sim$11\%. 
Model B (\pexrav), in spite of a significant improvement in statistics and a reasonable photon index, fails to treat the reprocessed components, i.e., the reflection component and the fluorescent lines, self-consistently. Therefore, in order to better unveil the physics of the obscuring matter surrounding the accreting SMBH of ESO 116-G018, more self-consistent and realistic models are needed.

\begingroup
\renewcommand*{\arraystretch}{1.5}
\begin{table*}
\centering
\caption{Summary of Best-Fits of XMM-Newton and NuSTAR Data using Different Models.}
\label{Table:best-fit}
\vspace{.1cm}
  \begin{tabular}{ccccccc}
       \hline
       \hline       
       Model&phenom&pexrav&MYTorus&MYTorus&MYTorus&borus02\\
       &&&(coupled)&(decoupled face on)&(decoupled edge on)&\\
       \hline
       $\chi^2$/d.o.f.&166/162&145/161&140/161&150/161&144/161&140/160\\
       $C_{Ins}$\footnote{$C_{Ins}$ = $C_{NuS/XMM}$ is the cross calibration between \NuSTAR\ and \XMM.}&0.99$_{-0.12}^{+0.13}$&1.03$_{-0.12}^{+0.13}$&1.06$_{-0.08}^{+0.08}$&1.01$_{-0.13}^{+0.14}$&1.07$_{-0.14}^{+0.15}$&1.07$_{-0.08}^{+0.08}$\\
       $\Gamma$&0.99$_{-0.18}^{+0.19}$&1.54$_{-0.19}^{+0.18}$&1.79$_{-0.12}^{+0.11}$&1.51$_{-0.11}^{+0.37}$&1.74$_{-0.34}^{+0.40}$&1.80$_{-0.06}^{+0.06}$\\
       norm\footnote{normalization of components in different models at 1\,keV in photons keV$^{-1}$\,cm$^{-2}$\,s$^{-1}$.} 10$^{-3}$&0.07$_{-0.03}^{+0.06}$&0.17$_{-0.09}^{+0.22}$&0.53$_{-0.34}^{+1.06}$&1.02$_{-0.47}^{+4.12}$&5.02$_{-5.02}^{+3.70}$&6.03$_{-0.23}^{+0.85}$\\
       R\footnote{Reflection scaling factor}&...&2.53$_{-1.16}^{+2.61}$&...&...&...&...\\
       N$\rm _{H,eq}$&...&...&5.31$_{-2.71}^{+2.34}$&...&...&...\\
       $\theta\rm _{tor}$\footnote{angle between the axis of the torus and the edge of torus in degree.}&...&...&...&...&...&81.69$_{-0.30}^{+0.88}$\\
       $f_c$\footnote{covering factor of the torus: $f_c$ = cos($\theta_{\rm Tor}$).}&...&...&...&...&...&0.14$_{-0.01}^{+0.01}$\\
       $\theta\rm _{obs}$&...&...&60.70$_{-0.40}^{+0.82}$&...&...&84.78$_{-0.88}^{+0.86}$\\
       A$_S$&...&...&4.12$_{-1.75}^{+2.38}$&0.76$_{-0.44}^{+0.47}$&0.45$_{-0.33}^{+0.88}$&...\\
       N$\rm _{H,Z}$\footnote{``line-of-sight'' column density in $10^{24}$\,cm$^{-2}$.}&0.62$_{-0.12}^{+0.12}$&0.88$_{-0.26}^{+0.32}$&...&1.46$_{-0.30}^{+0.57}$&2.58$_{-0.83}^{+1.21}$&2.60$_{-0.14}^{+0.16}$\\
       N$\rm _{H,S}$\footnote{``global average'' column density of the torus in $10^{24}$\,cm$^{-2}$.}&...&...&...&0.43$_{-0.13}^{+0.21}$&0.38$_{-0.08}^{+0.11}$&0.52$_{-0.06}^{+0.10}$\\
       $f_s$ 10$^{-2}$&10.69$_{-3.72}^{+5.94}$&1.44$_{-0.39}^{+0.40}$&1.67$_{-0.86}^{+0.91}$&0.77$_{-0.59}^{+0.91}$&0.27$_{-0.22}^{+0.94}$&0.16$_{-0.03}^{+0.03}$\\
       F$_{2-10}$\footnote{Flux between 2--10\,keV in $10^{-13}$\,erg\,cm$^{-2}$ s$^{-1}$.}&3.15$_{-0.09}^{+0.26}$&3.07$_{-0.21}^{+0.21}$&3.07$_{-0.38}^{+18.79}$&3.04$_{-0.28}^{+0.14}$&3.02$_{-0.33}^{+0.21}$&3.00$_{-0.68}^{+0.70}$\\
       F$_{10-40}$\footnote{Flux between 10--40\,keV in $10^{-12}$\,erg\,cm$^{-2}$ s$^{-1}$.}&3.33$_{-0.24}^{+0.28}$&3.45$_{-0.21}^{+0.40}$&3.49$_{-0.50}^{+1.38}$&3.34$_{-0.25}^{+0.33}$&3.47$_{-0.34}^{+0.25}$&3.41$_{-0.44}^{+0.35}$\\
       L$_{2-10}$\footnote{Intrinsic luminosity between 2--10\,keV in $10^{43}$\,erg\,s$^{-1}$, computed using `clumin' command.}&0.066$_{-0.004}^{+0.004}$&0.066$_{-0.008}^{+0.008}$&0.14$_{-0.02}^{+0.02}$&0.42$_{-0.04}^{+0.04}$&1.44$_{-0.17}^{+0.16}$&1.68$_{-0.19}^{+0.18}$\\
       L$_{10-40}$\footnote{Intrinsic luminosity between 10--40\,keV in $10^{43}$\,erg\,s$^{-1}$, computed using `clumin' command.}&0.25$_{-0.02}^{+0.02}$&0.11$_{-0.02}^{+0.01}$&0.16$_{-0.03}^{+0.02}$&0.74$_{-0.07}^{+0.07}$&1.81$_{-0.21}^{+0.21}$&1.86$_{-0.21}^{+0.21}$\\
       \hline
	\hline
	\vspace{0.02cm}
\end{tabular}
\end{table*}
\endgroup

\subsubsection{\MYTorus}\label{section:MYTorus}
\MYTorus\ models the intrinsic emission of an AGN reprocessed by obscuring matter with uniform density. The obscuring matter has a ``torus-like" structure with circular cross section, and the half-opening angle of the torus is fixed to $\theta\rm_{tor}$ = 60$^\circ$, i.e., the covering factor of the torus is fixed to $f_c$ = cos($\theta\rm_{tor}$) = 0.5. The angle between the observer ``line-of-sight'' and the torus axis (norm to the accretion disk), $\theta\rm_{obs}$, however, is a free parameter in \MYTorus\ and can vary in the range 0--90$^\circ$, where $\theta\rm_{obs}$ = 0$^\circ$ models a ``face on" scenario and $\theta\rm_{obs}$ = 90$^\circ$ models an ``edge on" scenario. Notably, the direct continuum will not intercept the circumnuclear matter when the inclination angle is less than $\theta\rm_{obs}$ = 60$^\circ$.

An advantage of the \MYTorus\ model is that the different components observed in the spectrum of an obscured AGN can be treated self-consistently. More in detail, the \MYTorus\ model is composed of three components: the direct continuum (MYTZ), the Compton-scattered component (MYTS) and the fluorescent emission-line component (MYTL). 

The direct continuum (MYTZ), which is also called zeroth-order continuum, is the ``line-of-sight'' observed continuum, i.e., the intrinsic X-ray continuum attenuated by the obscuring material in the torus. MYTZ is a multiplicative factor applied to the intrinsic continuum. In principle, the intrinsic continuum can be any continuum spectral shape: in our modeling we choose a power law one to be consistent with the scattered continuum and fluorescent emission-line components, which assume a power law incident continuum in \MYTorus. The direct continuum emission is an energy-dependent ``line-of-sight" quantity but is independent of the geometry of the torus.
 
The second component is the Compton-scattered continuum (MYTS), which is responsible for the ``Compton hump" observed at $\sim$10--40\,keV. The Compton-scattered continuum models those photons that are Compton-scattered into the ``line-of-sight" by the gas in the torus. The cutoff energy of the Compton-scattered component can vary in the range of $E\rm_T$ = [160--500]\,keV: we choose to fix this parameter to a standard $E\rm_T$ = 500\,keV value, since we verified that assuming a different cutoff energy does not significantly affect the other best-fit parameters. The covering factor of the \MYTorus\ model is fixed to be $f_c$ = 0.5: however, if the geometry of the torus differs significantly from the fixed \MYTorus\ value, or if there is a non-negligible time delay between the intrinsic continuum emission and the Compton-scattered continuum one, i.e., the central region is not compact and the intrinsic emission varies rapidly, the scattered component normalization can significantly differ from the main component one. To take these effects into account, the scattered continuum is multiplied by a constant, which we hereby define as $A_S$. 

Finally, the third component (MYTL) models the most prominent fluorescent emission-lines, i.e., the Fe K$\alpha$ and Fe K$\beta$ lines, at 6.4\,keV and 7.06\,keV, respectively. Analogously to $A_S$, the relative normalization between the fluorescent emission lines and the direct continuum is noted as $A_L$. In \XSPEC, $A_S$ and $A_L$ are implemented as two \texttt{constant}. Following previous works, the two relative normalizations are set to be equal, i.e., $A_S$ = $A_L$.   

In \XSPEC\ the \MYTorus\ model is described as follows:
\begin{equation}\label{eq:coupled}
\begin{aligned}
Model C =&constant_1*phabs*(MYTZ*zpowerlw\\
&+A_S*MYTS+A_L*MYTL\\
&+constant_2*zpowerlw)\\
\end{aligned}
\end{equation}

The \MYTorus\ model can be applied in two different configurations, named `coupled' and `decoupled' \citep{MYTorus2012}. We apply both configurations to the ESO 116-G018 spectrum: the analysis details and results are reported in the following sections. 

\subsubsection{\MYTorus\ in `coupled' configuration}\label{coupled}
In `coupled' mode, both the inclination angle and the column density are tied among the three components: MYTZ, MYTS, MYTL. In this configuration, the column density is the torus equatorial one, and the ``line-of-sight" column density is N$\rm _{H,l.o.s}$ = N$\rm _{H,eq}$ $\times$ $\rm(1-4\times cos(\theta_{obs})^2)^{1/2}$ \citep{MYTorus2009}.

The best-fit photon index for `coupled' \MYTorus\ model is $\Gamma$ = 1.79$_{-0.12}^{+0.11}$; the photon indices of the Compton-scattered continuum and of the fluorescent emission-lines component are tied to that of direct continuum. The equatorial column density is N$\rm_{H,eq}$ = 5.31$_{-2.71}^{+2.34}$ $\times$ $10^{24}$\,cm$^{-2}$ and the inclination angle is $\theta_{\rm obs}$ = 60.70$_{-0.40}^{+0.82}$$^\circ$, such that the ``line-of-sight" column densities are N$_{\rm H,Z}$ = 2.60$_{-1.36}^{+1.19}$ $\times$ $10^{24}$\,cm$^{-2}$. 

While the best-fit statistics of the `coupled' model is good ($\chi^2_\nu$ = 140/161 = 0.87), the geometrical scenario which the model presents is physically unlikely, since we measure $\theta_{\rm obs}$= 60.70$^\circ$ $\sim$60$^\circ$ = $\theta_{\rm tor}$, i.e., the AGN would be observed through the brink of the torus. Such a result also affects the reliability of the column density measurement, since N$_{\rm H,Z}$ is a parameter highly dependent on $\theta_{\rm obs}$, particularly when $\theta_{\rm obs}$ gets close to $\theta_{\rm tor}$. To further investigate the physical and geometrical properties of ESO 116-G018, we, therefore, try to apply \MYTorus\ in a different configuration, which allows one to disentangle the inclination angle and column density between the direct continuum and the reprocessed component.

\subsubsection{\MYTorus\ in `decoupled' configuration}
In `decoupled' configuration \citep{MYTorus2012}, the direct continuum and the Compton scattered component can in principle have different inclination angle and column density values. Since the direct continuum is a pure ``line-of-sight'' quality, which is independent on observation angle, the inclination angle of the direct continuum is fixed to $\theta_{\rm obs,Z}$ = 90$^\circ$, such that the column density of the direct continuum models the ``line-of-sight'' column density, N$_{\rm H,Z}$. The inclination angle of the Compton-scattered continuum and fluorescent lines is instead set to be either $\theta_{\rm obs,S,L}$ = 0$^\circ$ (face on) or $\theta_{\rm obs,S,L}$ = 90$^\circ$ (edge on), modeling a ``back-side'' reflection-dominated scenario or a ``near-side'' Compton scattered component-dominated scenario, respectively. In this configuration, the column density of the Compton scattered component and fluorescent emission-lines component parameterizes the ``global averaged'' column density of the torus, N$_{\rm H,S}$, which can significantly differ from the ``line-of-sight" column density in an inhomogeneous, patchy, torus \citep[see][for more details]{MYTorus2012}. Therefore, \MYTorus\ in `decoupled' configuration can be used to model a more realistic distribution of the obscuring material. We point out that the fluorescent emission lines component and Compton scattered component are still coupled since they are expected to be originated from the same process. 

\begin{figure*}
\begin{minipage}[b]{.5\textwidth}
\centering
\includegraphics[width=\textwidth]{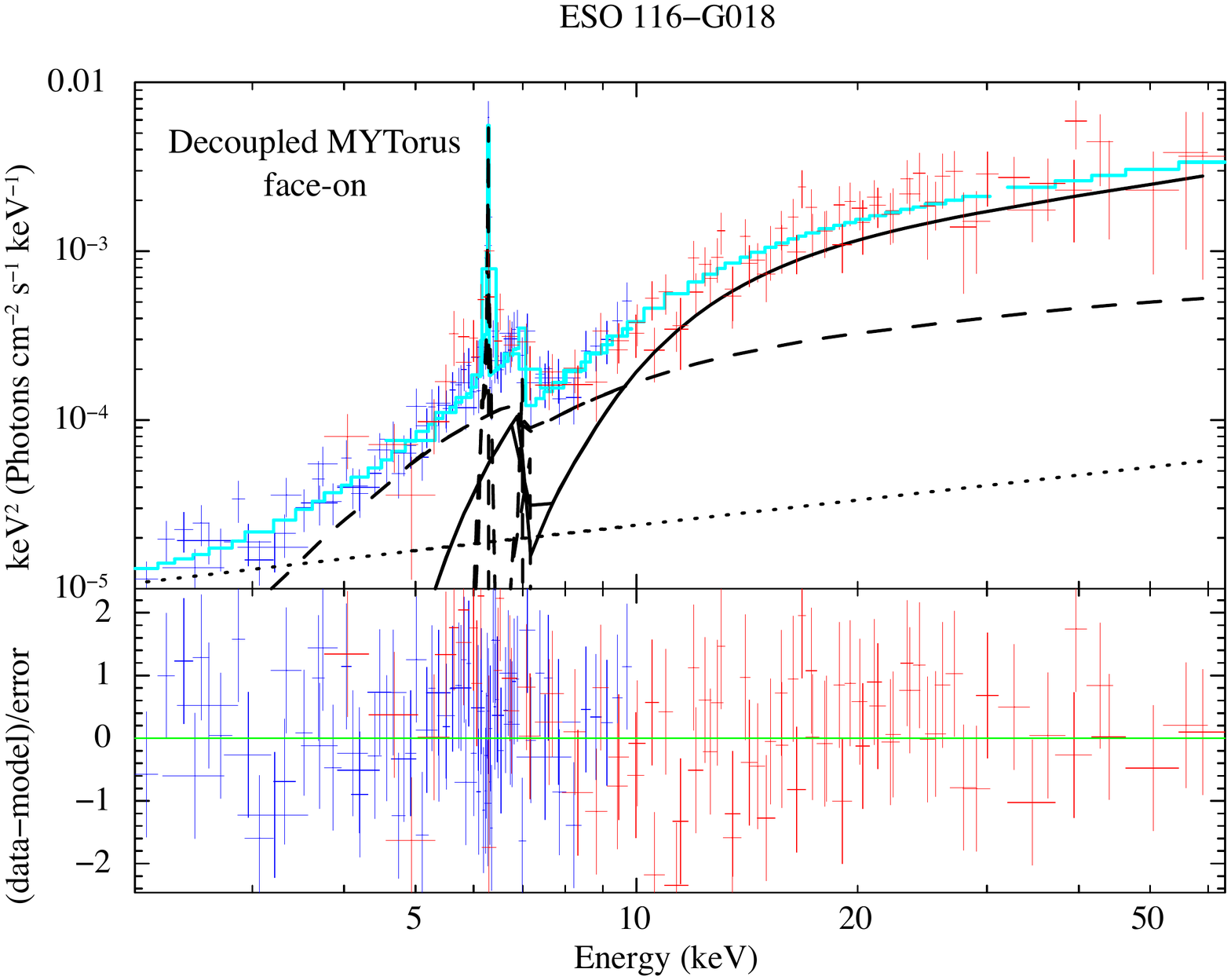}
\end{minipage}
\begin{minipage}[b]{.5\textwidth}
\centering
\includegraphics[width=\textwidth]{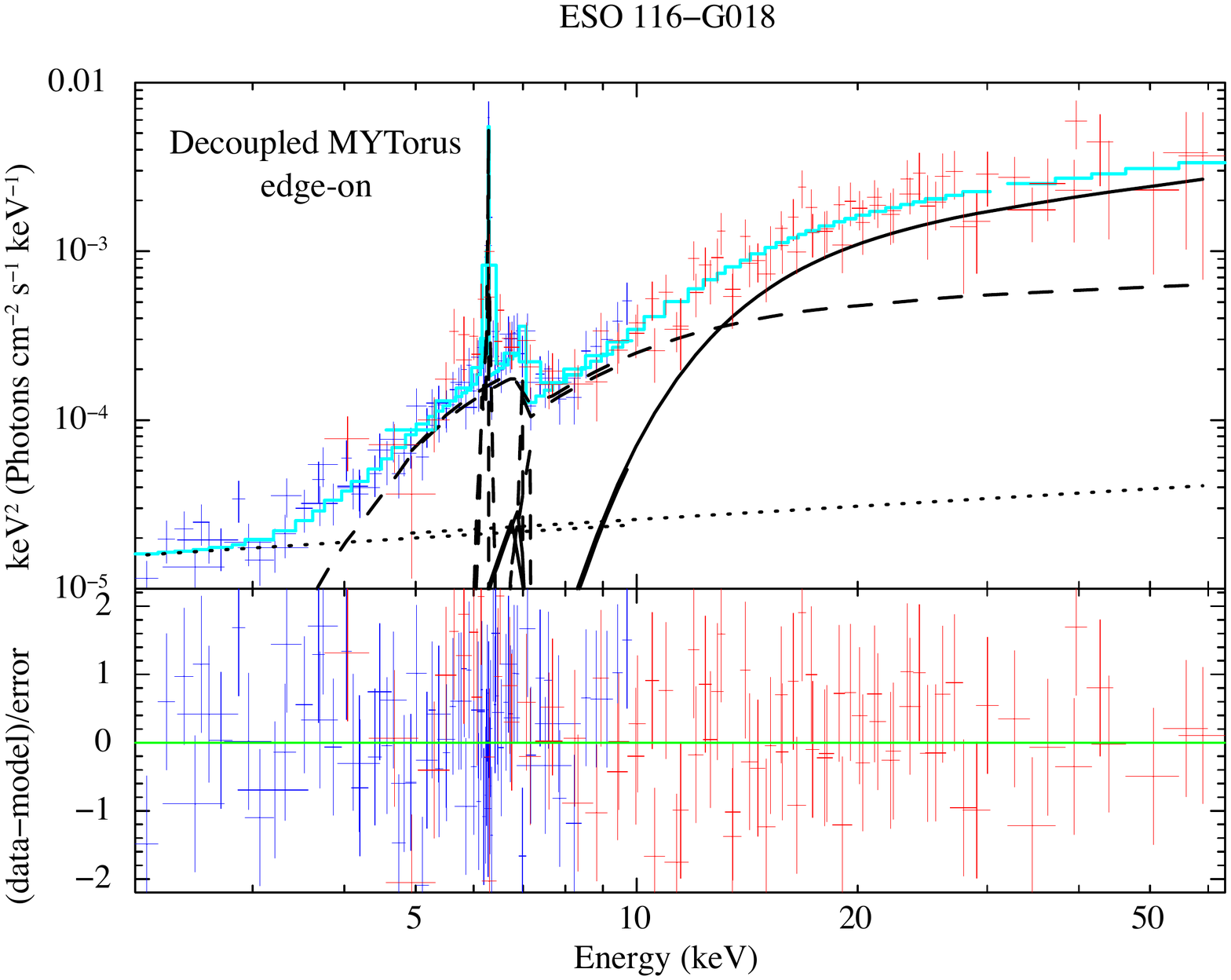}
\end{minipage}
\caption{Unfolded \XMM\ and \NuSTAR\ spectra of ESO 116-G018 fitted with `decoupled' \MYTorus\ ``face-on'' model (left panel) and `decoupled' \MYTorus\ ``edge-on'' model (right panel). The \XMM\ data are plotted in blue, while the \NuSTAR\ data are plotted in red. The best-fit models prediction is plotted as a cyan solid line. The single components of the model are plotted in black with different line styles, i.e., the absorbed intrinsic continuum as a solid line, the reprocessed component and Fe K$\alpha$ line as a dashed line and the scattered component as a dotted line.}
 \label{fig:decoupled}
 \end{figure*}

The best-fit photon indices are $\Gamma_{\rm \theta,S=0}$ = 1.51$_{-0.11}^{+0.37}$ and $\Gamma_{\rm \theta,S=90}$ = 1.74$_{-0.34}^{+0.40}$ for ``face on'' and ``edge on'' modes respectively. The ``line-of-sight" column densities are N$_{\rm H,Z,\theta,S=0}$ = 1.46$_{-0.30}^{+0.57}$ $\times$ $10^{24}$\,cm$^{-2}$ and N$_{\rm H,Z,\theta,S=90}$ = 2.58$_{-0.83}^{+1.21}$ $\times$ $10^{24}$\,cm$^{-2}$. The ``global average" column densities are N$_{\rm H,S,\theta,S=0}$ = 0.43$_{-0.13}^{+0.21}$ $\times$ 10$^{24}$\,cm$^{-2}$ and N$_{\rm H,S,\theta,S=90}$ = 0.38$_{-0.08}^{+0.11}$ $\times$ $10^{24}$\,cm$^{-2}$. The ``global average" column densities are $\sim$15\% and 29\% of the ``line-of-sight'' column densities for ``edge-on'' and ``face-on'' configurations, respectively, which suggests a patchy torus in ESO 116-G018. We present the unfolded \NuSTAR\ and \XMM\ spectra of ESO 116-G018 using the `decoupled' \MYTorus\ model in ``face on'' and ``edge on'' configuration in Fig.~\ref{fig:decoupled}.

In conclusion, the best-fit results of the \MYTorus\ model in both `coupled' \MYTorus\ model and `decoupled' \MYTorus\ model in ``edge on'' configuration confirm that ESO 116-G018 is a \textit{bona fide} CT-AGN at $>$3\,$\sigma$ confidence. While \MYTorus\ is effective in modeling the X-ray spectra of heavily obscured AGN, the `coupled' mode assumes a fixed torus opening angle ($\theta_{\rm Tor}$=60$^\circ$, i.e., a covering factor $f_c$ = cos $\theta_{\rm Tor}$ = 0.5), limiting the model to a single torus geometry and the `decoupled' mode fails to directly parameterize the geometrical properties of the obscuring material. To complement our analysis, we, therefore, model the ESO 116-G018 spectrum using the recently published \borus\ model \citep{Borus}, an updated version of the so-called \bntorus\ model \citep{BNtorus}. 

\subsubsection{BORUS02}\label{section:borus}
The model is composed of a reprocessed component (including the Compton scattered component and fluorescent lines) and an absorbed intrinsic continuum, described by a cut-off power-law, multiplied by a ``line-of-sight'' absorbing component, \texttt{zphabs}$\times$\texttt{cabs}. Although the $cabs$ model simply assumes a constant Compton scattering cross section equal to the Thomson cross section, which is in principle energy-dependent and only valid below $\sim$10\,keV, the difference between such a model and MYTZ is insignificant below $100$\,keV in our case, where N$\rm_{H,Z}$ $\sim2.6$ $\times$ $10^{24}$\,cm$^{-2}$ (more details are available in the \MYTorus\ manual\footnote{http://mytorus.com/mytorus-instructions.html}). In \borus\ the torus covering factor can vary in range of $f_c$ = [0.1--1], corresponding to a torus opening angle $\theta_{\rm Tor}$ = [0--84]$^\circ$. The observing angle ranges from $\theta_{\rm Tor}$ $\sim$[18--87]$^\circ$.

The \borus\ model is used in the following \XSPEC\ configuration:
\begin{equation}\label{eq:Borus}
\begin{aligned}
Model D =&constant_1*phabs*(borus+zphabs*cabs\\
&*cutoffpl+constant_2*cutoffpl) 
\end{aligned}
\end{equation}
where \textit{borus} is the reprocessed component in \borus.

\begin{figure}[htpb]
\centering
\includegraphics[width=.5\textwidth]{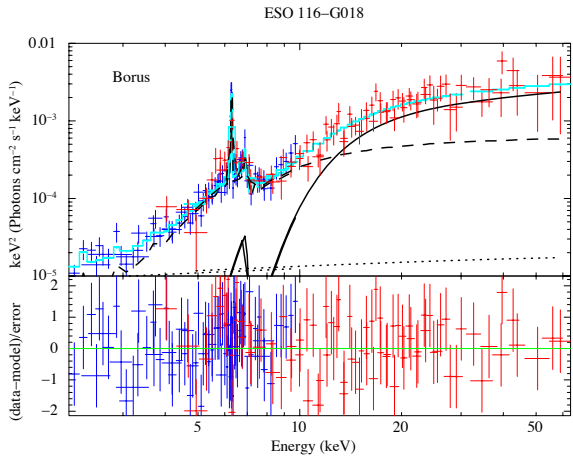} 
 \caption{Unfolded \XMM\ and \NuSTAR\ spectra of ESO 116-G018 fitted with the \borus\ model. The \XMM\ data are plotted in blue, while the \NuSTAR\ data are plotted in red. The best-fit models prediction is plotted as a cyan solid line. The single components of the model are plotted in black with different line styles, i.e., ``line-of-sight'' continuum as a solid line, reprocessed component as a dashed line and the scattered component as a dotted line.}
 \label{fig:Borus}
 \end{figure}

The best-fit photon index is $\Gamma$ = 1.80$_{-0.06}^{+0.06}$; the ``line-of-sight" column density is N$\rm _{H,Z}$ =  2.60$_{-0.14}^{+0.16}$ $\times$ $10^{24}$\,cm$^{-2}$; the column density of the torus is N$\rm _{H,S}$ = 0.52$_{-0.06}^{+0.10}$ $\times~10^{24}$\,cm$^{-2}$: in good agreement with the `decoupled' \MYTorus\ model in ``edge-on'' configuration. The half-opening angle of the torus is $\theta\rm _{tor}$ = 81.69$_{-0.30}^{+0.88}$$^\circ$, thus the torus covering factor is $f_c$ = cos($\theta\rm _{tor}$) = 0.14$_{-0.01}^{+0.01}$.
The inclination angle between the observer and the torus axis is $\theta\rm _{obs}$ = 84.78$_{-0.88}^{+0.86}$. The unfolded \NuSTAR\ and \XMM\ spectra of ESO 116-G018 fitted with \borus\ model is presented in Fig.~ \ref{fig:Borus}.

To conclude, the self-consistent, physically motivated models of \MYTorus\ and \borus\ give a better characterization of the X-ray spectrum than the phenomenological model and the \pexrav\ model, and confirm ESO 116-G018 to be a CT-AGN at $>$3\,$\sigma$ confidence. In addition, both the \MYTorus\ model in `decoupled' configuration and \borus\ display a significant difference between the ``line-of-sight" column density and the column density of the torus, suggesting a patchy distribution of the obscuring matter. The other physical properties of interest will be discussed in Section \ref{discussion}.

\subsection{Summary of the spectral analyzing results}\label{sec:summary}
In Section \ref{sec:pheno}, we first fitted the combined high-quality \XMM-\NuSTAR\ spectra of ESO 116-G018 using the phenomenological model A, finding that only an absorbed power-law is difficult to characterize the Compton hump in $\sim$10--40\,keV. We thus added a reflection component \pexrav\ to the model B to characterize the Compton hump: we obtained a significantly improved statistic, but the lack of self-consistency motivated us to explore more realistic models, i.e., \MYTorus\ and \borus, where the structure of the obscuring matter is a torus rather than a slab, like in \pexrav. In Section \ref{section:MYTorus}, we first tested the `coupled' \MYTorus\ model, which gave an unlikely geometrical scenario, i.e., the accreting SMBH would be observed through the brink of the torus. Therefore, we adopted the \MYTorus\ model configuration that assumes a more general geometry, by disentangling the direct continuum and the reprocessed component, and we then fitted the spectrum with the so-called `decoupled' \MYTorus\ model. This model gave a more reasonable result: however the obscuring material geometrical properties, i.e., the covering factor $f_c$ and observing angle $\theta_{\rm obs}$ cannot be directly derived using \MYTorus\ in `decoupled' configuration. Thus, we finally tested the recently published \borus\ model, which gives the best statistics and allows one to measure $f_c$ and  $\theta_{\rm obs}$.

Based on both the fit statistics and the reliability of the best-fit parameters, we believe that \borus\ provides the best-fit model for ESO 116-G018 in the 2--70\,keV band. In fact, while all the models used in our analysis have good fit statistics ($\chi^2_\nu$ $\sim$0.87--1.02), models other than \borus\ are limited by the assumptions required to use them or have notably different physical interpretations. For example, the \pexrav\ model and `coupled' \MYTorus\ model indicate that the spectrum is dominated by the reprocessed component, while the `decoupled' \MYTorus\ model and \borus\ model suggest that the direct continuum dominate the spectrum especially at energy $E$ $>$10\,keV. Such a discrepancy is also observed in other parameters, e.g., the intrinsic luminosity, which will be further discussed in Section \ref{sec:luminosity}. However, it is worth pointing out that in reprocessing-dominated models, such as \pexrav\, the reprocessed component is not obscured, while in `coupled' \MYTorus\ model, the reprocessed component is obscured by the dust and gas with the same column density as the ``line-of-sight'' one: both scenarios are unlikely to exactly characterize the real distribution of obscuring material, which has been shown to be clumpy, rather than uniformly distributed \citep[see, e.g.,][]{Krolik1988,Jaffe04,Tristram07,Nenkova08,Honig10,Stalevski12}. Indeed, the significant difference between the ``line-of-sight'' column density and the `global average' column density of the torus observed in the `decoupled' \MYTorus\ model and in the \borus\ model best-fits support a clumpy, patchy torus scenario for ESO 116-G018.

Regardless of the geometrical configuration of the obscuring material, both self-consistent, physically motivated models, \MYTorus\ and \borus, confirm the Compton-thickness of ESO 116-G018 at $>$3\,$\sigma$ confidence level. Furthermore, the \borus\ model provides excellent constraints on the geometrical properties of the obscuring torus: the covering factor is $f_c$ = [0.13--0.15] and we are observing the source ``edge-on''.

%
%
\section{Discussion and Conclusions}\label{discussion}
In this work, we report the results of the 2--70\,keV spectral analysis of ESO 116-G018, a nearby Seyfert 2 galaxy, observed quasi-simultaneously by \NuSTAR\ (45\,ks) and \XMM\ (58\,ks), and we establish for the first time that the source is a \textit{bona fide} Compton-thick AGN. As discussed in Section \ref{sec:summary}, the best-fit model for ESO 116-G018 is \borus\ \citep[][]{Borus}, which gives a ``line-of-sight'' column density of N$_{\rm H,Z}$ = [2.46--2.76] $\times$ $10^{24}$\,cm$^{-2}$. We also find that the best-fit results of `decoupled' \MYTorus\ model in ``edge on'' configuration \citep[][]{MYTorus2009} are in excellent agreement with those of \borus\ model. In the rest of the paper, we will use the \borus\ model best-fit results. 

\subsection{Comparison with previous results}\label{sec:compare}
ESO 116-G018 was found to be a CT-AGN candidate by \citet{marchesi2017APJ}, using a joint \cha-\textit{Swift}/BAT spectra fitted with the \MYTorus\ model in `coupled' configuration where the inclination angle is fixed to be $\theta\rm_{obs}$=90$^\circ$: they found a best-fit photon index, $\Gamma$ = 1.86$_{-0.48}^{+0.51}$, in good agreement with the one measured in this work  and a best-fit column density of N$\rm_H$ = 0.95$_{-0.40}^{+0.46}$ $\times$ $10^{24}$\,cm$^{-2}$, with $\chi^2$/d.o.f. = 17/14. Such a large uncertainty has been significantly improved by the high-quality data of the combined \NuSTAR-\XMM\ observations, which is potential the best combination of observatories to study the CT-AGNs in X-ray band.

\subsection{Equivalent width of the iron K$\alpha$ line}\label{sec:EW}
We are able to place strong constraints on the Fe K$\alpha$ line equivalent width (EW) of ESO 116-G018, due to the excellent count statistics provided by \NuSTAR\ and \XMM\ in the 5--8\,keV band, with a significant improvement with respect to \citet{marchesi2017APJ}. We use the task \texttt{eqwidth} in \XSPEC\ to measure the equivalent width EW$_{\rm phe}$ = 0.93$_{-0.15}^{+0.15}$\,keV and EW$_{\rm pex}$ = 0.85$_{-0.13}^{+0.15}$\,keV in models A and B, respectively. 

To measure the Fe K$\alpha$ line EW with \MYTorus\ we use the approach described in \citet{MYTorus2015}. We therefore first measure the continuum flux, without including the emission line, at E$_{\rm K\alpha}$ = 6.4\,keV. We then compute the flux of the fluorescent lines component in the energy range E = [0.95\,E$_{\rm K\alpha}$--1.05\,E$_{\rm K\alpha}$], i.e., between 6.08 and 6.72\,keV, rest-frame. EW is then computed by multiplying by (1 + \textit{z}) the ratio between the fluorescent line flux and the monochromatic continuum flux.
We obtain EW$_{\rm coupl}$ = 0.78$_{-0.13}^{+0.09}$\,keV, EW$_{\rm decoupl,\theta=0}$ = 0.90$_{-0.13}^{+0.10}$\,keV and EW$_{\rm decoupl,\theta=90}$ = 0.85$_{-0.12}^{+0.12}$\,keV. All \MYTorus\ equivalent width values are in good agreement with the ones obtained by phenomenological model and \pexrav\ model: furthermore, in all the models the measured Iron line EW value is typical of a CT-AGN \citep[$\sim$1\,keV; see, e.g., Fig.~8 in][]{MYTorus2009}.

\subsection{Intrinsic luminosity}
We report the 2--10\,keV and 10--40\,keV intrinsic luminosity in Table \ref{Table:best-fit}. Notably, the intrinsic luminosities derived from model B and the `coupled' \MYTorus\ model are $\sim$12--25 times smaller than those derived using \borus\ model the `decoupled' \MYTorus\ ``edge-on'' model. As discussed in Section \ref{sec:summary}, this is due to the fact that model B and `coupled' \MYTorus\ model are reprocessed-component-dominated, while \borus\ and the `decoupled' \MYTorus\ ``edge-on'' model are direct-continuum-dominated (at least at energies E$>$10\,keV). As already discussed in the previous sections, based on the statistics and reliability of parameters, we favor the \borus\ and \MYTorus\ decoupled ``edge-on'' solutions. The 2--10\,keV and 10--40\,keV intrinsic luminosity from the best-fit model are L$\rm_{int,2-10}$ = 1.68$_{-0.19}^{+0.18}$ $\times$ 10$^{43}$\,erg\,s$^{-1}$ and L$\rm_{int,10-40}$ = 1.86$_{-0.21}^{+0.21}$ $\times$ 10$^{43}$\,erg\,s$^{-1}$, respectively. This luminosity is compatible with the knee of the luminosity function of AGN in the local Universe \citep[][]{Ajello2012}, showing that ESO~116-G018 is an average-luminosity AGN.

The intrinsic X-ray luminosity can be derived indirectly from the luminosities measured at other wavelengths, such as the mid-infrared \citep[MIR, 3--20\,$\mu$m; see, e.g.,][]{Elvis1978}. The MIR flux of ESO 116-G018 is F$\rm_{12\mu m}$ = 0.175$_{-0.003}^{+0.002}$\,Jy \citep{wright10}, and the corresponding luminosity is L$\rm_{12\mu m}$ = 3.34$_{-0.05}^{+0.04}$ $\times$ 10$^{43}$\,erg\,s$^{-1}$. Applying the MIR-X-ray correlation in \citet{Asmus15} to the MIR luminosity, we obtain the 2--10\,keV luminosity to be L$\rm_{int,2-10,MIR}$ = 0.42$_{-0.10}^{+0.12}$ $\times$ 10$^{43}$\,erg\,s$^{-1}$. The 2--10\,keV intrinsic luminosity derived from MIR luminosity is slightly less than our best-fit one. This may be due to the fact that the covering factor of ESO 116-G018 is indeed small (see Section \ref{sec:cf}), which leads to a relative small infrared luminosity.

\subsection{Bolometric luminosity and mass of SMBH}\label{sec:luminosity}
The AGN bolometric luminosity is the measurement of the total AGN emission over the whole electromagnetic spectrum, and several bolometric corrections measurements to infer the bolometric luminosity from the X-ray one have been reported in the literature \citep[see,  e.g.,][]{Elvis1994,Marconi04,Lusso12,Brightman17}.
In Section \ref{sec:spectral}, we measured the intrinsic luminosity of ESO 116-G018 between 2--10\,keV which is L$_{\rm int, 2-10\,keV}$ = 1.68$_{-0.19}^{+0.18}$ $\times$ 10$^{43}$\,erg\,s$^{-1}$ using our best-fit \borus\ model. Applying the bolometric correction of \citet[][Equation 21]{Marconi04}, we obtain the bolometric luminosity of ESO 116-G018, which is L$\rm _{bol}$ = 2.99$_{-0.42}^{+0.42}$ $\times$ 10$^{44}$\,erg\,s$^{-1}$. 

Recently, \citet{Brightman17} measured the X-ray bolometric correction factors, $\kappa_{\rm Bol}$ $\equiv$L$\rm _{bol}$/L$_{\rm obs,8-24}$, where L$_{\rm obs,8-24}$ is the observed luminosity in 8--24\,keV, for CT AGNs. ESO 116-G018 8--24\,keV observed luminosity is L$_{\rm obs, 8-24}$ = 1.23$_{-0.12}^{+0.14}$ $\times$ 10$^{42}$\,erg\,s$^{-1}$, such that the bolometric luminosity from the prediction of \citet{Brightman17} is L$\rm _{bol}$ = 3.22$_{-2.11}^{+6.11}$ $\times$ 10$^{44}$\,erg\,s$^{-1}$, in excellent agreement with our results measured using the \citet{Marconi04} bolometric correction.

The Eddington ratio is a measurement of the SMBH accretion efficiency, and is defined as $\lambda_{\rm Edd}$ = $L_{\rm bol}/L_{\rm Edd}$, i.e., the ratio of bolometric luminosity, $L_{\rm bol}$, to the so-called Eddington luminosity, $L_{\rm Edd}$ = 4$\pi$GM$_{\rm BH}$m$_p$c/$\sigma_T$, where M$_{\rm BH}$ is the SMBH mass and m$_p$ is the mass of proton. Combining the bolometric luminosity, L$\rm _{bol}$, and the typical Eddington ratio of AGNs in local universe \citep[$\lambda$ $\sim$0.1; see, e.g.,][]{Marconi04}, one can estimate the mass of the center engine in ESO 116-G018, which is M$_{\rm BH}$ = $L_{\rm bol}$/(4$\pi$Gm$_p$c$\lambda_{\rm Edd}$/$\sigma_T$) and we obtain log\,(M$\rm_{BH}$/M${_\sun}$) $\sim$7.4, which is in agreement with the typical mass of SMBH: log\,(M$\rm_{BH}$/M${_\sun}$) $\sim$6.0--9.8 \citep[see, e.g.,][]{Woo02}.

\subsection{Covering factor}\label{sec:cf}
The self-consistency of the reprocessed components in \borus\ model provides one with the possibility to directly derive the geometrical properties of the obscuring material. In Section \ref{section:borus}, we measured the covering factor of the torus in ESO 116-G018 using \borus, and we found that a low-covering factor solution ($f_c$ = 0.14$_{-0.01}^{+0.01}$) is preferred. It is worth to note that this low covering factor could be explained both with a geometrically thin torus or with the fact that the torus is patchy, which is supported by the observed discrepancy between the ``line-of-sight'' column density and the `global average' column density of the torus.

The optical/UV disk emission re-processed by the torus in the infrared (IR) can also provide interesting constraints on the source geometry.
The ratio of the torus luminosity to the AGN luminosity can be thus be interpreted as the fraction of the sky obscured by the `torus-like' material, and the covering factor can be measured with the equation $f_c$ $\equiv$L$\rm_{tor}$/L$\rm_{AGN}$ = L$\rm_{IR}$/L$\rm_{Bol}$ \citep{Stalevski16}. \citet{Yamada13} measured the IR (8--1000\,$\mu$m) luminosity of ESO 116-G018, which is L$\rm_{IR}$ = 1.3 $\times$ 10$^{44}$\,erg\,s$^{-1}$, using the \facility{IRSA} \citep{IRAS} 12, 25, 60 and 100\,$\mu$m observations. Based on their measurement, and using the bolometric luminosity estimated from our best-fit model discussed in Section \ref{sec:luminosity}, the IR covering factor of the torus is $f_c$ = [0.38--0.51]. However, ESO 116-G018 is classified as a composite galaxy \citep{Yamada13}, i.e., the galaxy has both star-forming and AGN activity signatures, thus the IR luminosity is constituted of not only the AGN contribution but also a significant fraction of the polycyclic aromatic hydrocarbon (PAH) emission in star-forming process, suggesting that the covering factor of the torus of ESO 116-G018 should be smaller than that derived with the technique discussed above: the constraints on the geometrical property of the torus from the infrared study are thus in line with those obtained from our measurement in X-ray.

\subsection{Conclusion}
We find a significant difference between the best-fit results of the phenomenological model and \pexrav\ model, and the best-fit results of the self-consistent, physically motivated models. Such a difference is also found in NGC 1358 (Zhao et al. 2018 accepted).
In addition, \citet{Marchesi2018} re-examined the distribution of the column density of a sampled AGNs in local universe accompanied with \NuSTAR\ data, finding an overestimation of the column density than the one measured with \bat\ data probably due to the low-quality of the spectra. Since most of the AGNs are modeled with phenomenological model and non-self-consistent, physically motivated model in the previous analysis, the observed distribution of AGN will vary if self-consistent, physically motivated models are widely adopted. However, it is worth noting that the self-consistent, physically motivated models, i.e., \MYTorus\ and \borus, require high-quality data in broadband, therefore, as we show in this work and the work discussed above, the physically motivated model complemented with the combined \NuSTAR\ and \XMM\ analysis will be an ideal method to study the physics of heavily obscured AGNs.

%
%
\acknowledgements
The authors would like to thank L. Marcotulli for the help with the data reduction and C. Vignali for the helpful suggestions. X.Z., S.M. and M.A. acknowledge NASA funding under contract 80NSSC17K0635. NuSTAR is a project led by the California Institute of Technology (Caltech), managed by the Jet Propulsion Laboratory (JPL), and funded by the National Aeronautics and Space Administration (NASA). We thank the NuSTAR Operations, Software and Calibrations teams for support with these observations. This research has made use of the NuSTAR Data Analysis Software (NuSTARDAS) jointly developed by the ASI Science Data Center (ASDC, Italy) and the California Institute of Technology (USA). This research has made use of data and/or software provided by the High Energy Astrophysics Science Archive Research Center (HEASARC), which is a service of the Astrophysics Science Division at NASA/GSFC and the High Energy Astrophysics Division of the Smithsonian Astrophysical Observatory.

\bibliographystyle{aa}

\end{document}